\documentclass[11pt,twocolumn]{article}
\usepackage{graphics}

\usepackage{amsmath,amsfonts,amssymb,mathrsfs,graphicx}

\begin{document}
\title{Network of Perceptions}

\author{Fariel Shafee\footnote{fshafee@alum.mit.edu}\\
 Department of Physics\\ Princeton University\\
Princeton, NJ 08540\\ USA.}

\maketitle

\begin{abstract}
The role of perception in conscious behavior and decision-making is examined. The effect of spatial and temporal stochasticity in the acquisition of  beliefs is discussed. The idea of an agent as a locally strongly coupled group of states leads to the creation of energy minima in an interaction potential landscape. The interaction of such agent states and environment states acting at different levels of complexity and scale, subject to stochastically expressed interaction interfaces, may lead to asymmetry in perceptions. Agents possessing different perception related beliefs are then connected in a social network.
\end{abstract}

\section{Introduction}
Complexity and organization is a topic of current interest because of the many-faceted implications and applications in fields as diverse as biology (e.g. neural networks, see  Hopfield, 1995, and other models), social structure of elementary life forms such as bacteria, bird motion, organization of bees, and even the behavior of stock markets (Black and Scholes, 1973) and traffic control (see Helbing and Schreckenberg 1999).  Recent studies have also included elementary social behavior in human clusters (see Newman and Barabasi, 2006; Reichardt and White 2007).  The interdisciplinary nature of complexity makes it possible to cross boundaries of fields and include ideas from the natural sciences to explain conventional concepts of humanities. A model was suggested to encompass ideas of psychology, sociology, economics and physics to understand the organization of human societies (Shafee, 2004; Shafee 2009).  In this paper, we further investigate a specific aspect of that model, namely the role of perception and the inter-agent differences in perception, by analyzing their origin.

The concept of perception and related ideas of conscious thought processes has been a topic of heated debate.  Various vastly different models have been proposed (see Taylor 1999 for a comprehensive review). We borrow ideas from statistical physics, statistics, biology and psychology to investigate the emergence of perception and behavioral patterns arising from the role of perception within a social network.  Some interesting examples are also mentioned in the light of our model.

\section{The Interaction Hamiltonian}
An interaction based Hamiltonian was introduced in (Shafee ,2008; Shafee 2004) where the role of interactions among agents and between an agent and his local environment was seen to give rise to terms in the Hamiltonian function that governs time evolution. The model was based on the idea of spin glasses (Edward and Anderson, 1975; Sherrington and Kirkpatrick, 1975) where neighboring spins interact among themselves and also with a common environment.  The equation for the social interaction Hamiltonian from the point of view of an agent can be written as:
\begin{equation}
H_{potential} = H_{self}+H_{agent-agent}+H_{env}
\end{equation}

Here the respective Hamiltonian components have interaction terms of the form
$H_{self}= - J^{ab}_i s_i^a s_i^b$, $H_{env} = J^{a}_{ih} s_i^a . h^a$ and $H_{agent-agent} = J^{ab}_{ij} s^i_a s^j_b$

The $J$'s are the coupling constants, $i$ is the reference agent, $j$ is a neighboring agent, and $a$ is the label for a characteristic variable. The
$s$'s represent variable states within an agent and the $h$
's represent environment field components. The evolution of the interaction potential follows the law of steepest descent with respect to its variables. Hence, the system tries to reach energy minima (attractors) in the energy landscape.

The expression of each term is subject to blocking imposed upon one term by another interaction term (mutual exclusiveness of some pairs of interactions), so that is the blocked part of the interaction, say  $J_{ij}^D S_i^D s_j^D$ involving the pair of agents $i,j$ and variable $D$, due to the interaction, say, $J_{ij}^C S_i^D s_j^D$.  The origin of blocking and expression is the need for sharing of strongly coupled components that are barred from being independent in order to maintain a potential minimum.

\section{Degree of Complexity}
\subsection{Degree of Complexity and Localization}

 At the most elementary level, the possibilities of stable or semi-stable structures are specific in terms of components, and finite discrete groups exist. In physical science we know that a quark can pair up with two other quarks in elementary hadronic particles and the stability of such structures depend on almost similar quarks that are dissimilar in the sense that they have different colors interacting locally, so that the total color is zero.  The matches and the complementary qualities required to produce the stability due to interactions, make it possible only to have a few permutations of quark types to produce these elementary particles.  A proton is identical with another proton, and symmetrization of wave-functions are required at the quantum level to account for the uncertainty in identifying protons in a compound system.

However, the interaction range of the strong force at the quark level makes the stable structures extremely localized and shielded from outside interaction.  A free quark is never seen in reality.  A quark and an anti-quark form a strong bond and the lines of strong force must end in an anti-color, with no leakage of information (see Shafee, 2007 e.g. for a detailed discussion).

As the complexity level goes up, less complex components become packed within a more complex structures, which is spatially larger than the individual components (eg, a human is larger than the proton where the three quarks are very localized within the entity of the proton and can rarely be exchanged with quarks outside the proton).

\subsection{Degree of Complexity and Fuzziness}

As complexity gets higher, more weakly shielded interactions come into play, and diverse categories of components can be seen within the same structure.  For example, although a proton contains only various types of quarks held by the strong force, at the next level of complexity an atom contains both hadrons and leptons, and the force holding these structures together have a larger interaction range, and offer less shielding.  So, the components of one structure may be within the interaction range of another structure, and these interactions may give rise to further complex structures.  For example, a group of atoms can be held within a molecule where the orbits of the electrons overlap.

However, the number of protons in an atom is somewhat flexible, with each number giving rise to a distinct type of element with different properties of interaction with other elements.

Hence, we can relate degree of fuzziness with a level of hierarchy. Similarly, a monkey and a human may both be primates, but the brain size of a monkey is smaller than that of a human.  So, although the identities of two monkeys may be distinct due to genetic variations, the variation is not known {\em a priori}.  But the set of differences between a monkey and a human may derive largely from the cognitive advancement of a human because of the existence of more complex brains and mutations. The complexity of the brain gives rise to a larger degree of choices and cognitive processes. Hence, even though two human beings differ in terms of personality, they share some traits because of the increased level of complexity relative to a monkey.

Although two human beings can be seen as members of the same level of complexity, they are not exactly the same. The number of cells in a human is only approximate, while the number of quarks within a proton is exact. However, at the level of organs, two human beings share virtually the same number of organs that are quite similar.  The slight variation comes from environmental and hormonal effects, genetic makeup etc.  Hence the stability of each organ within an environment depends on the other organs, the gene that maps the organ etc.  The range of viable organs is dependent on a threshold and not an exact number.

The survival of a gene, which is the blueprint of a human being, depends on the produced agent being within the permitted range of variables for survival, given the environment and the basic laws of nature required for the interactions that couple the components within the agent.  Two human beings may not be equally adept in a specific aptitude, and one may be more robust with respect to a certain variable.  However, the statistical fluctuations of the many units of a gene make the degree of robustness with respect to different variables distributed, hence obstructing the existence of a super-human.  The complementary robust variables make it favorable to have collaborative interactions among agents, creating social networks.

\section{Macro-States and Reorganization}
In the construction of the model the concept of variables in different states and the concept of interacting states are used.  On the macroscopic scale, the state of a variable may derive from the reorganization of the components of the system, which again may be due to interaction with another  system. The two possible states of hemoglobin can be a good simple example.  The final state (or shape) of a hemoglobin depends on how many oxygen molecules are bound to its sites, so that the interaction with oxygen causes a hemoglobin to be in a more adaptive and efficient state. As energetically favorable connections or reorganizations are sought to form a favorably bound system, it is possible for one state even to split into smaller states and dissipate because of energy kinetics.

In a broader sense, multiple systems can interact with one system, subject to the interaction range and available contact surface for interaction and binding, and the reorganized final state would depend on the totality of the types of interacting states, and the duration of interaction.

The reorganization energy is a measure of stiffness of the state.  The rate of transfer of energy between two
imperfectly matched bound states signify the duration of binding needed to bring about a reorganization. 

A similar scenario can be seen when a macroscopic detector reorganizes itself to express a quantum state after quantum measurement ( see von Neumann, 1932 eg).

\section{Perception of the Global Environment}
Perception is the transfer of information between an environment  and an agent or between agents by means of interaction. The transfer of information is brought about by disturbing or reorganizing the perception organ temporarily by means of interaction with an environment system, and then transferring the reorganization information to permanent or semi-permanent states within the brain.  This transfer is brought about by means of the couplings of the organs with the brain.  The perturbed organ of  perception reverts to the original state by its connections with other components of the agent that also interact with the environment, so that energy is transferred from the environment to the agent in order to reset the perturbed organ to the original state with respect to the other more constant structures it is attached to. An excited reorganized perception organ falls back to the original state after distributing the excitation signals to other connected organs. Hence, the organ's state depends on its current interaction with the environment, and also its couplings or interactions with the agent's own component systems.

The stability of memory states within the brain depends on the complexity and vastness of neurons within the brain, so that the slight change of the weight of a connection is accommodated within the macroscopic aggregate structure with respect to its connections with other organs (Buszaki, 2006). However, the large number of internal connections make these minuscule weight changes produce significantly different behavioral patterns.  Hence, the perpetuation of the identity of the agent is related to its ability to revert macroscopic organ connections to original states, together with the ability of the intricately connected brain to permanently modify its connections to map the input signals.  However, though the weights of the neurons change to reflect agent-environment interactions to preserve interaction information locally within the agent, the topology of the internal connections forming in the first few years by means of genetic commands and the effects of the local environment, remain stiff and protected from environment-inputs, so that the agent's  personality and some preferences remain stiff ( Rauch et al 2005 for an example of a genetically determined trait).  The identity of the agent derives from the interactions between these two parts, and, hence, is a function of genetically fixed traits continued in the background of
modified information and beliefs acquired from agent-environment interfaces.

\subsection{Perception and Dimensionality}
The many degrees of freedom within the environment may be seen as an independent aspect, in the same manner phase spaces are created.
Due to slight variations of the genetic makeup of perceptions, various agents, although connected to the same
environment, may be able to perceive the environment slightly differently. They will have slightly
different ideas about the world.  This is analogous to being connected to a shadow (projection) of a higher dimensional world where all agents are connected, although interacting
with one shadow  may change the entire higher dimensional world and hence alters another shadow as well. So two
agents connected to slightly different shadows, which cannot perceive each other's worlds completely, cause
changes to the other agent's world by interacting with his own world.  Though the first agent will conceive
his own interactions with his own world completely rational, he might find the interaction of the second agent
with respect to the first world irrational, even though the second agent himself will find interactions with
his own world perfectly rational.

\subsection{Emergence of Beliefs }
The emergence of the different projections can be derived by using stochasticity and coarse graining. The specific value of an interacting state can be recorded by using perception organs.  Hence, a perception-type interaction provides information about a certain state of the environment, that can be translated into information about corresponding correlated states or possible reactions to specific actions (interactions that realign a state or reorganize an environment state into separate states).

However, the need for couplings of organs and the need for components of organs to act in unison to transfer a signal or piece of information necessitates the transfer of information to be at the scale of the perception organ and connections.  Hence, a threshold of intensity is needed for a retina cell to fire a visual signal to the brain (see Ratliff, 1974), and the human ear can transfer audio signals within the range of 20 to 20000 Herz. Again, the finite width of the action potential in a neuron, necessary to maintain  a time-scale and the measure of simultaneity (see Buszaki, 2006 for a discussion) imposes a coarse graining scale in resolution, and the refractory period of a neuron originating from the necessity of several subcomponents (ion channels e.g.) to act in unison for the detection of a signal imposes a refractory period when no signal can be recorded, causing information to be lost.

 The complexity of the system makes it possible to add diverse categories of signals from different perception organs sharing a set of coupled organs.  Hence, the sharing requires the precision of one category to impose a cost on another category.

 The stored information, which transfers into a belief, is thus a function of these scales and interconnections. As mentioned above, the fuzziness associated with two agents at a complexity level causes the exact scales and couplings to vary slightly within the {\em range} of survival, and hence, due to the added complexity of the number of internal connections and couplings, the same environmental state may get translated into separate belief states within two agents.

Again, the dimension of the complex system requires that it connects with environmental states of similar dimensions within the same scale of the expressed organs because of the interaction ranges and shielding.  This scaled ``window" into the environment causes each agent to be connected to a local environment window, although the global environment causes the local window to change. The stochastic fluctuation within two {\em local} windows, where two agents are connected, makes them perceive a global variable differently, giving rise to conflicting beliefs about the same state.

\section{Perception of Rules as Cues from Past Interactions}
The idea of simultaneity of two events and temporal overlaps of two events are recorded within the brain by means of the finite width of the action potential of neurons, which forms a time-scale for experience sequences (see Buszaki, 2006). When two environment states are recorded simultaneously, a correlation is formed within the brain between the two states by the nature of the mapping mechanism.  Again, temporal cues are able to correlate with one event preceding another (Raaijmakers, 1979; Raaijmakers and Shiffrin, 1980).  So, when a piece of information related with the same state is fed again to the brain, it causes the previously stored signal of the same image to be activated, exciting an entire cue of correlated images(Raaijmakers , 1979). The perception of a current state activates the images of a series of states that are correlated with the fed image by means of past association.

\subsection{Cued Perception and Temporal Stochasticity}
The chaotic, complex and periodic dynamics differ in the degree of uncertainty associated with determining future orbits. The difference between static, orbital, complex and chaotic systems can be attributed to the fact that while the behavior of a chaotic system at a certain time is independent of its stored historical sequences, and the behavior of a static or orbital system is determined, a complex system behaves in a complicated manner, taking into account its history to a certain degree (see Palis, 2002 for a review of chaotic, complex and periodic systems and the history of the theory).
The degree of complexity or chaotic behavior may derive from placing this {\it window} of perception locally within the environment (within an agent's identity) in a vast network where different levels of such complexity arise.

The simple laws of nature arising from the interaction forces give relations for aggregate behavior of systems in different scales.  For example, gravity makes a planet rotate around the sun in a periodic manner defined by the mass of the sun and the mass of the planet, and the distance between them.  How the masses are reorganizing within the planet at a small scale does not affect this aggregate periodicity.  However, when a slice of the planet is considered, it is connected with a larger part of the planet, but in the same time, within it, smaller scales of interaction are seen that depend on the exact position of each atom. The degree of history dependence of a system is dependent on the storage of history, and,  therefore, interface interaction of system and environment.  A system with an inadequate storage mechanism  depends completely on the forces of the immediate environment state.  The behavior may become periodic when the mean field of the environment is large and homogenous, and chaotic when the environment fluctuates and appears to be random in the locality because of the inability of the system to collect random data-points and relate it with simple rules at a larger scale.  The stochasticity arising from the loss of information at each interaction also leads to the accuracy of construction of a  correct history.

If an agent can locally store past interactions and use cues to past resultant states, when a certain state is presented, the idea of global futures and hence future global states at the scale of the interaction states derive from the accuracy of the intercepted rules at that scale within the larger environment.  Although the agent connects with its local environment, in multiple interactions regarding a certain states in different localities and different times act as data-points in time and space regarding rules at that scale.  Hence the system's dependence on history is based on it being able to accurately fit data-points to a larger graph.  This is similar to the idea of binning so that the size of bins may produce different graphs, which approach the continuous graph when bins are small.

The dependence of history also is contingent upon rules existing among each simple unit of the vast degree of freedom of the environment, so that further rules can be derived for states existing in different scales, but comprising these simple units.  The loss of information due the large degrees of freedom is thus compensated by the stability of states existing at larger scales with properties and approximately similar rules deriving from the average interactions of the components.  Hence, the finer information, such as the exact location of a single unit may not be expressed in the rules displayed by the larger states, but may be a smooth function on``totals",``averages", ``differences" etc. of the elementary properties, giving rise to macroscopic parameters expressed in the macroscopic scale.

\subsection{Local cued Perceptions and Global Costs}

As we have commented above, a principal difference between intelligent human ensembles and inanimate passive ones is that in the latter the values of the variables are defined uniquely, but in the human systems the agents may perceive own or others' values to be different from an agent-independent assessment. So, subsequent decisions and choices, i.e. the dynamical changes, may depend on the extent of the distortion of individual perception.
The use of imagination to anticipate the non-immediate future enables human agents to take decisions which are not necessarily the best instantaneous goals, but are usually better in the longer run.
A short term increase in the interaction potential can take place if virtual future interactions are introduced in the perceived Hamiltonian. This local increase in interaction potential due to  interaction of existing terms with future virtual terms can be seen as second order couplings, where virtual, perceived states are coupled with existing state-wise interactions.

\subsection{Perceived Hamiltonian and Actual Hamiltonian and Cued Connecting Interactions}
  The  continuation of an identity in time depended on an agent's interactions or choices of  interactions that were based on the current states, which also contained information about past interactions.  Cues acted as temporal correlations of interactions and states with past interactions and states. Hence, perception in the form of memories are able to evoke virtual perceived states.

When a certain state is local to an agent, it can thus evoke a cue to a possible future state and interaction, related with an interaction potential (may involve emotions).  This type of couplings between real and virtual states (which may be called second and higher order couplings in a perturbative series) via an intermediate state (which may not produce the ``present optimal connection" but a cue to a future optimal connection with a current high weight) may include virtual terms and interactions in the agent's Hamiltonian.  This type of interactions can be represented diagrammatically as follows:
The agent may decide to interact with a present non-optimal state by adding these higher order interactions and virtual terms in his  perceived Hamiltonian, which is also the interaction energy perceived by the agent within his own reference frame.  The effects of these local perceived terms may be expressed in the global actual interaction landscape as a result of the agent's actions dependent on these perceived local interactions.

\subsection{Local Cost and Perception}

The addition of perceived terms in the agent's perceived interaction potential may cause the agent-environment interactions to be non-optimal at a certain time. Hence, a cost term is introduced. Cost may be formulated in our context in the following way.

Let us for brevity define an optimal interacting pair $[A]$ as $w_A J^i_{aa'}s_i^a t^{a'}$, where $t$ may be either agent variable $s$ or environment variable $h$,  and the generic symbol $A$ stands for the optimal pair $aa'$, and $w_A$ is the probabilistic weight for this pair. Let $B_{AB}$ be the blocking matrix which shows the weakening of a pair $B$ because of the formation of the pair $A$, when there is some kind of mutual exclusion. So, the actual optimum Hamiltonian is

\begin{equation}
H_{act-opt} = \sum_A ( [A] - \sum_B B_{AB} [B] )
\end{equation}

If, in addition to the actual states, we also have additional virtual states (perceived from past associations in real experience, skill, training etc.), so that we have a generalized optimal pair labels $X,Y$ that include both $a$  states and virtual $p$ states, then the perceived Hamiltonian

\begin{equation}
H_{per} = \sum_X \left( [X] - \sum_Y B_{XY} [Y] \right)
\end{equation}

If, however, we now truncate the sum to remove the perceived states $p$, and express the Hamiltonian in terms of only the actual states, we shall no longer get the optimal Hamiltonian

\begin{equation}
H_{act-nonopt}= H_{per} - \sum_P [P]
\end{equation}
Here the set $P$ involve at least one state in the perceived group.

Now the cost may be defined as

\begin{equation}
C= H_{act-opt} - H_{act-nonopt}
\end{equation}

This cost will create an abrupt increase in the total interaction energy due to the local non-optimal interaction because of the existence of the virtual interaction term based on future expectation.
The idea of blocking interactions was discussed in detail in (Shafee, 2009).

\begin{figure}[ht!]
\begin{center}
\includegraphics[width=6cm]{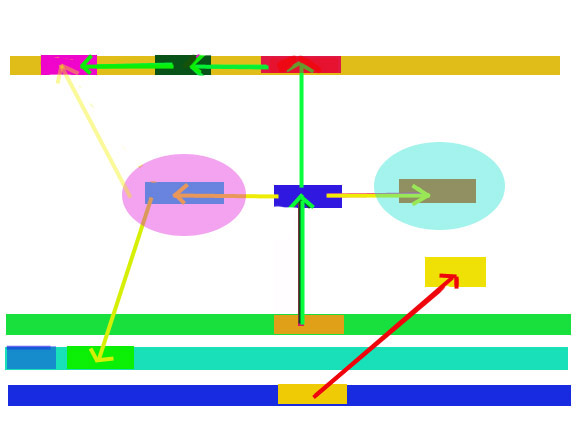}
\end{center}
\caption{\label{fig1}Connections between perceived and actual terms in the present and the future. Top (yellow) array contains virtual terms, bottom (blue) array is a present optimal agent-array, without virtual terms, the next (light blue) array is a future array, and the next (light green) array is the present optimal array including virtual terms. Light green arrows indicate interactions in $H_{perc}$, where the dark blue square connected with the orange square in the present optimal $H_{perc}$ is an agent-environment connection so that the connection causes the environment to be one of the two possible future states (within circles) depending on environmental stochasticity.  If the light blue state is created, it couples with the magenta perception state and increases the strength of the cue, $\alpha$. $H_{act}$ is the part of $H_{perc}$ without the perception terms, and is not optimal.  If the perception states were not present, the optimal current array would have been the dark blue one, corresponding to $H_{act-opt}$.  The difference between $H_{act}$ and $H_{act-opt}$ gives the cost of including virtual perception terms, with the probability of a more optimal future Hamiltonian}

\end{figure}

\subsection{Perceived Hamiltonian and the Affordability of Costs}

In (Shafee, 2009; Shafee, 2007), agents were formulated as semi-closed systems that gained their local stability in an interaction energy landscape because of the stable couplings within the agent locally, i.e. through self-interaction.  While the internal couplings of the agent's variables, giving rise to an internal interaction Hamiltonian $H_{self}$, sustained the agent's identity within the environment, the expression of the agent within the environment occurred because of the interactions between the agent variables and the environment and the variables of other agents.  Higher order identities in the form of groups originated because of these inter-agent and agent-environment interactions.

However, it was also discussed in (Shafee, 2009), how a critical threshold of $H_{self}$ is required in order to maintain the stability of the agent.  This value, $H_{crit}$, retains the symbiotic couplings.  If the value is higher than $H_{crit}$, there is not enough couplings or interconnections among the agent components, and the symbiotic stability is lost.  The agent components and variables instead get connected with environmental variables, or get transformed into non-optimal or smaller components that cannot be retrieved because of the many degrees of freedom of the environment that these components get connected with.  Hence, once the agent's internal coupling is above a threshold, the agent is out of the game irrevocably. $H_{crit}$ as the cutoff of $H_{self}$ exists in the global field where the agent is a locally strongly coupled system, and $H_{self}$ is the interaction Hamiltonian of self-interaction within that highly coupled localized part.  When costs  occur within the agent's own perceived Hamiltonian, that takes into account expressed variables and also virtual future terms, the limit of costs incurred is balanced by the threshold of $H_{self}$. As a result, the effect of the cost cannot push $H_{self}$ above $H_{crit}$.  Therefore, if the incurring of a cost leads to the separation of two terms in $H_{self}$, that need to be strongly coupled in order to keep $H_{self}$ above (weaker bound) $H_{crit}$, that cost cannot be incurred. Similarly, if adding virtual terms in the perceived Hamiltonian pushes $H_{self}$ at the present time above $H_{crit}$, those virtual terms are not allowed in the agent's perceived Hamiltonian.
So, in terms of present and future possible components, the perceived Hamiltonian can be modified from $H_{act}$  as
\begin{equation}
H_{perc} =H'_{act}(C(H_{virt}))+ H_{virt}
\end{equation}

$H_{virt}$ is the interaction potential due to the interactions of the virtual term.  The first term on RHS, $H'$ is the modified $H_{act}$ because of the costs  imposed by the terms in $H_{virt}$ (both the first order and second order changes as described above, due to the changed degrees of expression of various terms in $H_{act}$ when the virtual interaction pair is introduced) , and also because of the second order pairwise couplings created between possible  terms of $H'_{act}$ and $H_{virt}$ so that
\begin{equation}
H_{perc} \leq H_{act}
\end{equation}
with the further constraint that $H'_{act}(C(H_{virt}))  \leq H_{crit}(local\space identity)$
The last inequality follows the definition of a local identity as being a local minimum in the interaction Hamiltonian so that highly coupled local variables interact with neighboring variables from the environment to produce a minimum that resists an increase in the energy.
When agents are dispersed in the locality and the effect of collaborating strong identities can be neglected, individuals have random matching and mismatching components scattered in the environment.  The strength from the bond of inter-agent interactions is, therefore, much weaker than the strong symbiotic couplings of the components within the agent. Under these conditions, $H_{crit}(local\space identity) = H_{crit}(agent)$, and the agent, in this zero'th order, behaves rationally by interacting with the environment and by trading with others to optimize his own selfish needs.
	Hence, Eq. 6 is, in most cases constrained by $H'_{act}(C(H_{virt}))\le H_{crit}$ for the agent, and the agent's perception and actions are limited by his own preservation.  However, once $H_
{act} \le H_{crit}$ for the agent, the agent's own coupled identity is lost irreversibly.
However, if the constitution of the group of agents is changed in terms of total matches and total number of agents, the zero'th order identity is distorted and modified.  An extreme example is discussed in (Shafee 2008).

\section{Long-term and Short-term Goals}
A perception Hamiltonian involving only short term or temporally local virtual terms causes the interactions to be locally optimal in time, often causing sudden changes in the environment states not foreseen (Barnett et al 2005).  Again, the calculation of long term satisfaction often comes with the cost of blocking local interactions, and causing local discomfort and sometimes lack of knowledge of immediate future states. Extreme cases can be: leading an acutely uncomfortable life to save every penny for the rainy day.  However, such actions also come with the cost of calculations of many approximate interactions in a series, and the payoff for the sufferings may never take place because of sudden forks in the time-line caused by unforeseen changes due to the introduction of unseen and uncalculated variables introduced to the locality by means of its global connection.

\begin{figure}[ht!]
\begin{center}
\includegraphics[width=6cm]{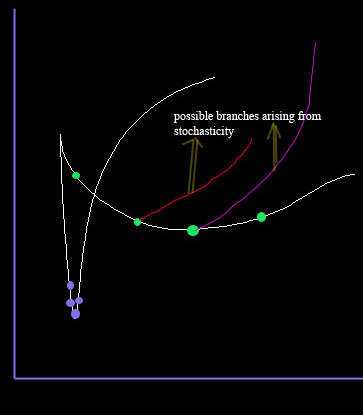}
\caption{\label{fig2}The effect of long term and short term goals based on cued possible events (circled. Green circles are long term planning with overall stability and clue circles imply short term pleasure with possible non-optimal scenarios in the long term.)}
\end{center}
\end{figure}

\section{Calculation of Terms by Recombination vs. Cues}
In the simplest case, a virtual future term is evoked in the Hamiltonian by means of a cue. The cue automatically conjures a less than optimal interaction, based on past experience or belief.  The inclusion of these perceived virtual term allows for the continuation of the identity over a temporal spread, as was discussed before.

The {\bf equation 2 and 3} include a term $\alpha$, which is the strength of the virtual term.  If the virtual interaction appears as a direct cue, strengthened by emotional attachment, the value of $\alpha$ is large, and the connection is made instantaneously by reflex.

However, a different way of including virtual terms is by adding a desired interaction having a large coupling with an existing state, and finding a series of cues to add present environment states to interact with present available states.  The search for a local environment state that would connect with a present agent state to create a path starting at a present agent-environment interaction and ending at the desired future virtual state connected with the first state requires carrying out a search in all cued sequences.  The search consists of connecting with activating various images of states within the brain, and using neurons shared with perception for the case of exhaustive searches.  The process thus blocks (see Shafee 2008 for details of blocking) processing actual states that the agent is connected with at present.  So, the search of distant histories blocks recent histories from being recorded, and reduces the current state's dependence on recent history.  As a result, the agent may be ill-prepared for short term future, and may in turn incur high costs.

Also, as one cue is recombined with another, stochastic uncertainties from the separate processes add up, and the possibility of ``actually" obtaining the final environment state is also reduced.

Again, in the extreme end, the connection of a present state with a virtual state without any cued path results in a low interaction potential of the perceived Hamiltonian (see Shafee,2004; Shafee, 2009).  As a result, the agent may be satisfied within himself with no connection with reality or the future.

\section{Game of Fuzzy Logic}

In this section, we study the effect of fuzzy logic derived from perception within a social network. A similar basic game using agents in a network interacting with different axioms was proposed (Shafee, 2002). We extend the concept by introducing detailed mechanism with different balancing factors causing various behavioral patterns, including herd mechanisms, lynching, and at the other end, disorganized chaos, and forced alignment (brain washing).

The agents are given beliefs that may have degrees of overlap.  The beliefs make up rules and cues that dictate an agent's interactions and states.  The modes of formation of a belief (interactions with other agents or interactions with an environment), and the credibility lent to a piece of belief come into play. Within the game, the agents try to optimize their move by obtaining many data points to form a picture, since, as described previously, a cue is only probabilistic and dependent on the number of similar past  interactions.

However, the price of a piece of information depends on the number of interactions needed to be made and the ``blocks" imposed by one such interaction on another interaction.  The localization of an agent also makes it difficult to gain simultaneous data points regarding a state's interaction at another geographical locality in order to form patterns of ``global rules" needed to find the global dependence on the local environment.

Communication with another agent may transfer information about one agent's interaction and perceived states to another agent.  This form of communication is dependent on the common overlap in perception mechanism between the two agents, conflicts of interest (competing preferences attached to one environment) (For details of matches between two agents, see Shafee 2009; Shafee, 2004), and also biases held by an agent due to his own past experience irrelevant to the second agent (so, the correlation of one agent's own local experience with the piece of information supplied, not known to the second agent). However, communication with agents in different localities comes with the advantage of reducing the number of blocks (exclusive pairings) in the agent's own interaction time-line.

The credibility of a piece of information obtained by an agent can thus be a function of the `` known" match between the agent and the provider (say $\mu$) and the frequency of the same information received from unknown and unverified agents, assuming random matches as approximation (say, $f$).
The credibility of such information is balanced by the number of experiences involving the ``belief" experienced by the agent in first person, and the number of interaction data-points pertaining to  the belief within the agent, acquired by his own interactions.  Therefore, the high cost of obtaining a belief on his own is balanced by the frequency and credibility of ``cheap" and imperfect (from another agent's perception) beliefs obtained by communication with other agents.

Specific related cases may be apparent threats to an agent leading to herd mechanism and mechanisms like lynching.  Two other extreme scenarios are: 1. agents are forced to believe in the same axioms, which may be contradictory to their own genetic biases, and 2. agents communicating in small groups within isolated localities and suddenly exposed to one another.

We study the first example in detail.

Herd Mechanism by Threat: When a piece of information is associated with an imminent threat, the cost of not believing the information comes with the risk of the threat being correct. The credibility of the information or belief depends on the following:

1. The existence of no or few conflicting beliefs in the following form
agent-agent belief $\rightarrow$ $s_i^a.s_j^a$ $\rightarrow$ $s_i^b.s_j^{b'}$, where $s_j^{b'}$ is not optimal with $s_j^b$.

2. The credibility of the source of information in the form of trust or matching.  This may derive from genetic kinship, membership in the same highly weighted interest group or the source's affiliation with a reputed institution.

3. The frequency of the information received. If already $n$ members of a cluster is ``converted", the $n+1$- th member has a higher probability of being converted.  Such phenomena have been studied in the case of spread of infectious diseases or the spreading of rumor (see Kawachia et al 2008).  The alignment along the direction of the fad has been studied .

4. The duration of time between the possibility of the hypothetical threat maturing and the time required to personally obtain the information by interacting with the environment or the state in question.

5.  The degree of threat in the personal interaction.  For example, if there is a rumor that a certain snake-bite is deadly, an agent would be reluctant to personally get bitten to verify the information.  The observation of data comes with the risk of trusting the data.  Having another person getting bitten by the risk in front of the first person has less risk factor for the first person, but is unlikely to take place in most cases.

Hence, instead of using the simple function used for observing phase transition in possible two-state existence such as in the spread of disease or rumors in a connected network (see e.g. Kawachia 2008), or the creation of social networks based on homogenous systems (Newman and Barbasi, 2006; Reichardt and White, 2007), the balancing factors are introduced for the case when herd mechanism involves risk factors for an agent, and is limited to simple alignment along the group with respect to fashion trends.

If a contradictory cue exists within the agent such that the belief evokes $s_j^b$ with strength $\alpha$ and the cue evokes $s_j^{b'}$ with strength $\alpha'$, then the same stimulant will give rise to a mixed state of cues, $\alpha s_i^b+\alpha' s_i^{b'}$.  Both the evoked non-overlapping images would give rise to virtual terms, and the degree of cost associated with each virtual interaction would be present in the perceived Hamiltonian.  This may be equated with situations of ambivalence.

The other terms affecting credibility described above can be used to find a relation between the strength of
an axiom obtained from the environment, as opposed to one obtained from agent-agent connection. The conformity of  the particular agent $i$ with other agents $j \neq i$ is given on the average by $\sum_j J_{ij}^{aa'} s_i^a s_j^{a'}/N'$, where $N'$ is the number of other agents who cover the same traits as $i$. This may be normalized by dividing by the self interaction which has complete harmony to get $P$, the normalized measure of accord with a random agent
\begin{equation}
P= \sum_j J_{ij}^{aa'} s_i^a s_j^{a'}/(N'J_i. s_i.s_i)
\end{equation}
The probability of accepting other agents' axioms will, therefore, be a product of $P$ times a stochastic measure $\mu$ that corrects for the value for small, unreliable samples (small N'), and a time factor  T that induces a sense of perceived urgency, as determined by the importance of the information or the belief in terms of survival or great damage or benefit (a high risk or reward demands immediate acceptance of other agents' belief, without the risk of self/direct interaction with the environment to form first hand belief):

\begin{equation}
Q=  P\mu(N') T(t-t')
\end{equation}
where $t$ is the perceived time required for direct interaction, and $t'$ is the perceived time at hand till the peak of the $H$ associated with the coupling of the agent's variable $s_i^a$ with the environment is reached. The function $T$ may be modeled as per importance of genetic and other factors $G$ and scaled by the magnitude $R$ of the risk/reward, e.g. we may have in the simplest case
\begin{equation}
T=1-GR(t'-t)
\end{equation}
with $GRt'~1$, which ensures that when there is plenty of time at hand $T~0$, or when the credibility of other agents is small, the agent might prefer to interact with the environment directly, and if the credibility is high, and the time constraint is severe, the agent would prefer to trust other agents and reassess his orientation $s_i^a$.

In the special case when distinct mutually  mismatching sub-groups exist within a group, a bias in circulating the information may create distinct subgroups that may come together.

\section{Conclusion}

We have modeled the origin of individuality from the point of view of stochasticity experienced by an agent both spatially and temporally when connected with a global time-line.  This results in variation in perceptions.  The creation of beliefs and cues locally within the agent shows that the difference in beliefs causes the agents to adjust their actions accordingly and may often put an agent in a temporarily uncomfortable or unfavorable position.  The effect of connecting agents with slightly different perceptions also gives interesting features.  Further detailed analysis of group behavior subject to individual perception and belief differences, as mentioned in the last section, may be worth investigating.
\section*{Acknowledgement}

The author would like to thank Prof. Douglas White and Prof. Bertrand Roehner for their continued encouragement related to my work.

\section*{Reference}

Black F,  Scholes M. The Pricing of Options and Corporate Liabilities. J Pol Econ 1973; 81 (3): 637-654.
\\

Barnett TP, Adam JC,  Lettenmaier DP. Potential impacts of a warming climate on water availability in snow-dominated regions. Nature 2005; 438 (7066): 303–309.
\\

Buzsaki G. Rhythms of the Brain. New York (NY): Oxford Univ Press; 2006.
\\

Edwards S, Anderson P. Theory of spin glasses. J Phys F 1975; 5: 965-974.
\\

Helbing D, Schreckenberg M. Cellular automata simulating experimental properties of traffic flows. Phys Rev E  1999; 59: R2505-R2508.
\\

Hopfield  JJ, Herz AVM.  Rapid local synchronization of action potentials: toward computation with coupled integrate-and-fire neurons. Proc Natl Acad Sci 1995; 92: 6655.
\\

Kawachia K, Sekia CM, Yoshidab H. A rumor transmission model with various contact interactions. J Theor Bio  2008; 253(1):  55-60.
\\

Newman MEJ, Barabási AL, Watts DJ. The Structure and Dynamics of Networks. Princeton (NJ): Princeton Univ Press; 2006.
\\

Neumann VJ.  Mathematical Foundations of Quantum Mechanics. Princeton (NJ): Princeton Univ
 Press;
  1932. p. 195-197.
\\

Palis J. Chaotic and Complex Systems. Curr Sci 2002; 82(4):  403-406.
\\

Raaijmakers JGW.  Retrieval from Long Term Store: A General Theory and Mathematical Models.  Unpublished Doctoral Dissertation, Univ Nijmegen, The Netherlands 1979.
\\

Raaijmakers JGW and Shiffrin RM. SAM: A Theory of Probabilistic Search in Associative Memor. In: Bower GH, editor. The Psych of Learning and Motivation: Advances in Research and Theory. New York: Academic Press; 1980. pp. 207-262.
\\

Rauch SL, et. al. Orbitofrontal thickness, retention of fear extinction, and extraversion.  Neuroreport 2005; 16(17):1909-12.
\\

Ratliff F. Studies on Excitation and Inhibition in the Retina: A collection of Papers from the Laboratories of H. Keffer Hartline. New York: Rockefeller Univ Press; 1974.
\\

Reichardt J, White D. Role Models for Complex Networks. Proc of the Nat Acad of Sci  USA 2007; submitted.
\\

Shafee F. A Spin Glass Model of Human Logic Systems. xxx.lanl.gov, arXiv:nlin/0211013; 2002.
\\

Shafee F. Chaos and Annealing in Social networks. xxx.lanl.gov, arXiv:cond-mat/0401191; 2004.
\\

Shafee F. Oligo-parametric Hierarchical Structure of Complex System. NeuroQuantology J 2007 5(1):85-99.
\\

Shafee F. Identity and Interaction of Agents in Social Networks. submitted; 2009.
\\
\\

Sherrington D   and Kirkpatrick SK, Solvable Model of Spin Glass.  Phys Rev Lett 1975; 35: 1792-1796.
\\

Taylor JG. The Race for Consciousness. Cambridge ( MA) : MIT Press; 1999.

\end{document}